\documentclass{aa}
\usepackage[colorlinks,citecolor=blue,linkcolor=blue,urlcolor=blue]{hyperref}

\title{Long-term optical spectroscopic variations in blazar 3C~454.3}
\titlerunning{Optical spectroscopic variations in blazar 3C~454.3}
\author{
Krzysztof~Nalewajko\inst{\ref{inst-ncac}}
\and
Alok~C.~Gupta\inst{\ref{inst-aries}}
\and
Mai~Liao\inst{\ref{inst-shang},\ref{inst-beij}}
\and
Krzysztof~Hryniewicz\inst{\ref{inst-ncac}}
\and
Maitrayee~Gupta\inst{\ref{inst-ncac}}
\and
Minfeng~Gu\inst{\ref{inst-shang}}
}
\authorrunning{Nalewajko {et~al.}}
\institute{
Nicolaus Copernicus Astronomical Center, Polish Academy of Sciences, Bartycka 18, 00-716 Warsaw, Poland\\
\email{knalew@camk.edu.pl}
\label{inst-ncac}
\and
Aryabhatta Research Institute of Observational Sciences (ARIES), Manora Peak, Nainital 263002, India
\label{inst-aries}
\and
Key Laboratory for Research in Galaxies and Cosmology, Shanghai Astronomical Observatory, Chinese Academy of Sciences, Shanghai 200030, China
\label{inst-shang}
\and
{University of Chinese Academy of Sciences, 19A Yuquanlu, Beijing 100049, China}
\label{inst-beij}
}

\abstract {}
{Characterisation of the long-term variations in the broad line region in a luminous blazar, where Comptonisation of broad-line emission within a relativistic jet is the standard scenario for production of $\gamma$-ray emission that dominates the spectral energy distribution.}
{We analysed ten years of optical spectroscopic data from the Steward Observatory for the blazar 3C~454.3, as well as $\gamma$-ray data from the Fermi Large Area Telescope (LAT). The optical spectra are dominated by a highly variable non-thermal synchrotron continuum with a prominent Mg~II broad emission line. The line flux was obtained by spectral decomposition including significant contribution from the Fe~II pseudo-continuum. Three methods were used to characterise variations in the line flux: (1) stacking of the continuum-subtracted spectra, (2) subtracting the running mean light curves calculated for different timescales, and (3) evaluating potential time delays via the discrete correlation function (DCF).}
{Despite very large variations in  the $\gamma$-ray and optical continua, the line flux changes only moderately ($< 0.1\;{\rm dex}$). The data suggest that the line flux responds to a dramatic change in the blazar activity from a very high state in 2010 to a deep low state in 2012. Two interpretations are possible: either the line flux is anti-correlated with the continuum or the increase in the line luminosity is delayed by $\sim 600\;{\rm days}$. If this time delay results from the reverberation of poorly constrained accretion disc emission in both the broad-line region (BLR) and the synchrotron emitting blazar zone within a relativistic jet, we would obtain natural estimates for the BLR radius $R_{\rm BLR,MgII} \gtrsim 0.28\;{\rm pc}$ and for the supermassive black hole mass $M_{\rm SMBH} \sim 8.5\times 10^8 M_\odot$.
We did not identify additional examples of short-term `flares' of the line flux, in addition to the previously reported case observed in 2010.}
{}


\begin{document}

\maketitle

\section{Introduction}

The cores of active galaxies exhibit a complex mixture of energetic physical phenomena (for review see \citealt{2016ARA&A..54..725M}, and references therein). Supermassive black holes ($M_{\rm SMBH} \sim 10^9M_\odot$) accrete surrounding gas, often forming radiatively efficient accretion discs that produce intense radiation fields (quasar emission; $L_{\rm disc} \sim 10^{46}\;{\rm erg\,s^{-1}}$) that ionise the surrounding gas, resulting in emission lines broadened by rapid gas motions ($v > 1000\;{\rm km\,s^{-1}}$). In some cases, magnetic flux accumulated on the black hole produces a pair of relativistic collimated outflows called jets that produce non-thermal radiation beams (blazar emission), the luminosity of which can be greatly enhanced (by a factor of $\sim 10^4$ to $L_{\rm jet} \sim 10^{48}\;{\rm erg\,s^{-1}}$) by special-relativistic effects for observers located along one of the jets. A class of active galaxies called flat spectrum radio quasars (FSRQs) combines the characteristics of blazars, where  broad non-thermal spectral components dominate the radio, infrared, optical, X-ray, and gamma-ray bands,  and those of quasars,  where thermal emission of accretion disc peak  in the rest-frame UV, and broad emission lines (BELs) produced in so-called broad line region (BLR) contribute to the rest-frame optical and/or UV.
The BELs constitute an important source of soft radiation that strongly interacts with relativistic jets, with the potential of producing the bulk of observed $\gamma$-ray emission by inverse Compton (IC) scattering off ultra-relativistic electrons \citep{1994ApJ...421..153S}, and of  absorbing a part of the $\gamma$-ray spectrum in the process of photon-photon pair production \citep{Bla95}. The understanding of  the BLR properties, especially   its geometry and dynamics, while essential for making a self-consistent picture of FSRQs, is still rather basic \citep{Tav08,2011A&A...525L...8C,2012arXiv1209.2291T,2015MNRAS.449..431J}.

A primary example of an FSRQ of exceptional gamma-ray luminosity (up to $10^{50}\;{\rm erg\,s^{-1}}$ \citealt{2011ApJ...733L..26A}) and large-amplitude time variability is 3C~454.3 ($z = 0.859$). It has been a target of multiple dedicated multi-wavelength observational campaigns \citep{1993ApJ...407L..41H, 2006A&A...445L...1F, 2006A&A...449L..21P, 2006A&A...453..817V, 2006A&A...456..911G, 2007A&A...473..819R, 2008A&A...491..755R, 2010ApJ...715..362J}, especially in the context of high-energy gamma-ray observations by AGILE \citep{2008ApJ...676L..13V, 2009ApJ...690.1018V, 2009ApJ...707.1115D, 2010ApJ...712..405V, 2011ApJ...736L..38V} and by Fermi-LAT \citep{2009ApJ...699..817A, 2009ApJ...697L..81B, 2010ApJ...721.1383A, 2011MNRAS.410..368B, 2011A&A...534A..87R, 2012ApJ...758...72W, 2012AJ....143...23G, 2013ApJ...773..147J, 2017MNRAS.464.2046K, 2017MNRAS.472..788G}.

As one of the brightest gamma-ray blazars of the Fermi era, 3C~454.3 was selected for regular optical spectroscopic monitoring by the Steward Observatory \citep{2009arXiv0912.3621S}.
Based on a part of this dataset (2008-2011), a flare-like variability of the Mg~II broad emission line flux in 2010 November (${\rm MJD} \sim 55512$) was reported by \cite{2013ApJ...763L..36L}.
However, independent optical spectroscopic survey by the SMARTS Consortium over a very similar period but with relatively sparse cadence found variations in the Mg~II flux at the level of $2\sigma$, in addition to more significant variations in the H$\gamma$ line \citep{2013ApJ...779..100I,2015ApJ...804....7I}. Here, we analyse ten years of spectroscopic data from Steward Observatory comprising   564 observations, an unprecedented dataset in terms of long-term regular and frequent monitoring. We find a hint (statistical significance of $\sim 1.5\sigma$) of modulation in the Mg~II line luminosity on a timescale of about two years in response to large-amplitude variability of the optical and gamma-ray continua that were not   noticed by previous studies. The line luminosity is either anti-correlated with the optical continuum luminosity or delayed by $\sim 300$ source-frame days.
This would be only the second case of reverberation mapping in a blazar after  the case of 3C~273 (PG~1226+023; $z = 0.1583$) included in the sample studied by \cite{2000ApJ...533..631K}, with a recent analysis based on the Steward survey \citep{2019ApJ...876...49Z}.

\begin{figure*}
\centering
\includegraphics[width=0.98\textwidth]{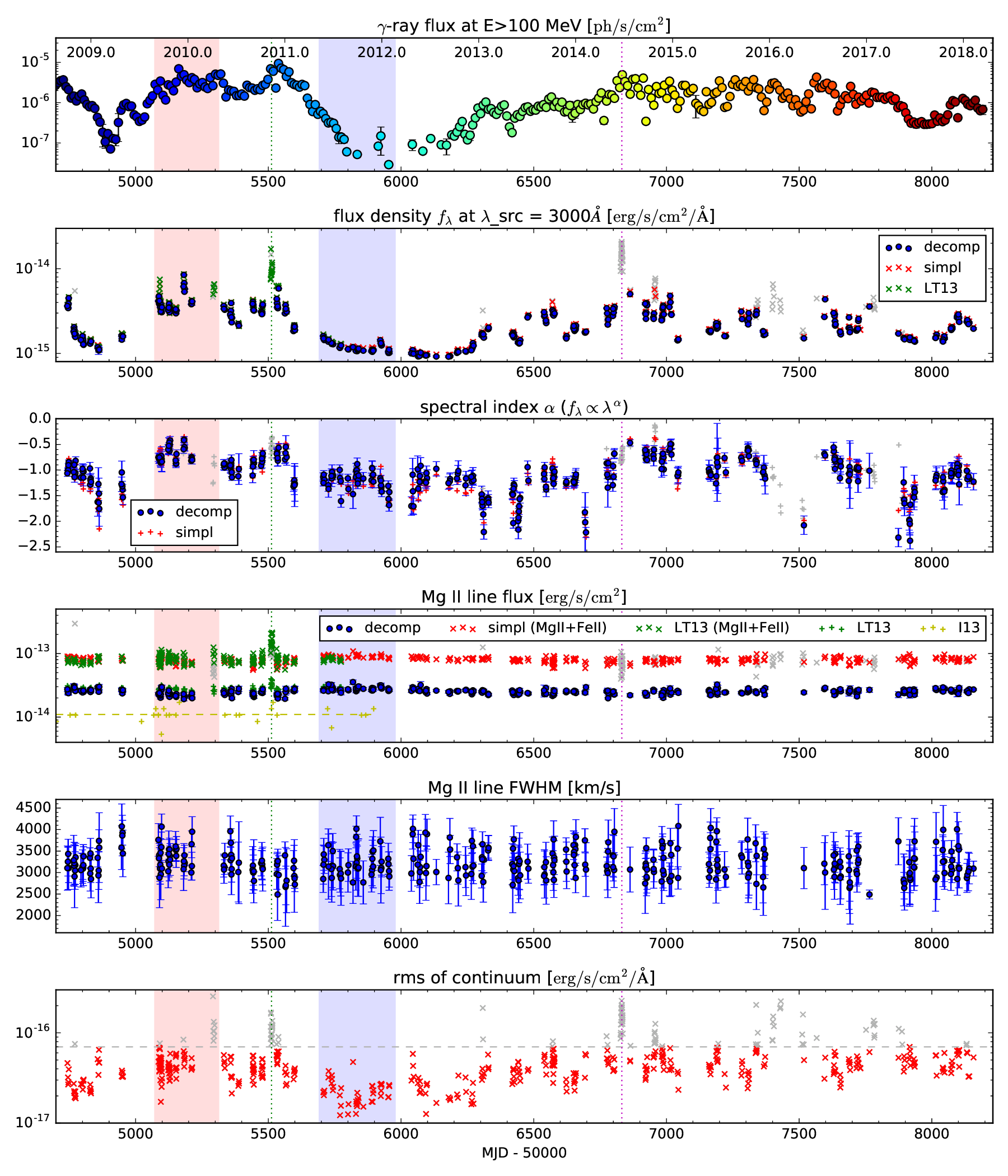}
\caption{Light curves of blazar 3C~454.3. From the top, the first panel shows the $\gamma$-ray photon flux for photon energies above $100\;{\rm MeV}$ obtained from the Fermi-LAT data in 10-day bins; the second panel shows the optical continuum at source-frame $3000\AA$, obtained independently using two analysis methods (spectral decomposition,  blue circles; simplified power-law fit,  red/grey crosses); the third panel shows the corresponding values of the spectral index $\alpha$ such that $f_\lambda \propto \lambda^\alpha$; the fourth panel shows estimates of the spectrally integrated Mg II line flux, either separated from the Fe II continuum (decomposition method) or not (simplified method); the fifth panel shows the source-frame FWHM width of Mg II line obtained in the decomposition method and corrected for instrumental broadening; the sixth panel shows the rms of continuum subtracted spectra calculated over the continuum fitting spectral windows, with the limiting value of $7\times 10^{-17}\;{\rm erg/s/cm^2/}${\AA} indicated with the dashed line. Results reported in \cite{2013ApJ...763L..36L} (Steward) and \cite{2013ApJ...779..100I} (SMARTS) are indicated in the second and fourth panels. Periods of interest for spectral stacking are marked with vertical stripes or lines.}
\label{fig_lc}
\end{figure*}

\section{Data}

\subsection{Steward Observatory}

Regular optical photometric and spectropolarimetric observations of a sample of gamma-ray bright blazars are performed by the team of the University of Arizona Steward Observatory as support programme for the Fermi Gamma-Ray Space Telescope\footnote{\url{http://james.as.arizona.edu/~psmith/Fermi}}. Moderate-resolution ($R \sim 300-1000$) optical spectra were obtained in the spectral range of $\lambda_{\rm obs} = (4000 - 7600)${\AA} with the SPOL spectropolarimeter and calibrated to the $V$ magnitudes of nearby standard stars. Technical details of the instrument and data reduction procedures are described in \cite{2009arXiv0912.3621S}.

We analysed all 564 photometrically calibrated spectroscopic observations over the period of ten years (2008-2018; ${\rm MJD} = 54743 - 58306$).
{The observed spectra were corrected for Galactic extinction using the map of \cite{1998ApJ...500..525S} and the extinction curve from \cite{1989ApJ...345..245C} with the following parameters: $A(V) = 0.33\,{\rm mag}$ and $E(B-V) = 0.105\;{\rm mag}$ with $R_V = 3.1$.}

The main spectral feature seen in all spectra is the Mg~II broad emission line ($\lambda_{\rm src} = 2800${\AA} redshifted to $\lambda_{\rm obs} = 5200${\AA}. It is imposed on a continuum consisting of non-thermal synchrotron emission component of the blazar and of thermal accretion disc emission component of the quasar, as well as an Fe~II pseudo-continuum. As the primary analysis method, we performed spectral decomposition of the Mg~II line complex following the approach of \cite{2011ApJS..194...45S}. The Mg~II emission was modelled with up to three Gaussian components, with the Fe~II template adopted after \citep{2001ApJS..134....1V}, and the continuum was fitted within the source-frame spectral windows of $(2200-2700)${\AA} and $(2900-3090)${\AA}.
The centroids of the three Gaussians were treated as free parameters, and their values show stochastic variations without a systematic long-term trend. Sorting the Gaussians according to their centroids (low, medium, high), the median source-frame centroid values are $2764{\rm\AA}$, $2797{\rm\AA}$, and $2820{\rm\AA}$, with the standard deviations of $14{\rm\AA}$, $3{\rm\AA}$, and $12{\rm\AA}$, respectively.
The observed full width at half maximum ${\rm FWHM}_{\rm obs}$ of the Mg~II line was calculated from the superposed profiles of the fitted Gaussian components, and it was corrected for instrumental line broadening using ${\rm FWHM}_{\rm int} = ({\rm FWHM}_{\rm obs}^2 - {\rm FWHM}_{\rm instr}^2)^{1/2}$ with the adopted source-frame instrumental resolution of ${\rm FWHM}_{\rm instr} \simeq 1150\;{\rm km\,s^{-1}}$.
In order to estimate observational flux uncertainty, we calculated the rms value of the spectra over the source-frame spectral window of $(3000-3100)${\AA} after subtracting a second-order polynomial function. To estimate uncertainties of the measured spectral parameters, we generated samples of 100 mock spectra by adding a random Gaussian noise to the observed spectrum using the flux density error and calculating the rms of the obtained distributions of measured parameters.
A more accurate theoretical template for the Fe~II pseudo-continuum, with independent Fe~II multiplets taken from \cite{2015ApJS..221...35K} and \cite{2019MNRAS.484.3180P}, was fitted to the mean line profile; however, it was not used to model individual observations.

We also perform a simplified spectral analysis in order to verify that the inferred modulation of the Mg~II line flux is not affected by uncertainties in the Fe~II pseudo-continuum template and its separation from both the Mg~II line or from the actual non-thermal continuum. In this approach, the continuum is evaluated as a power law fitted over two narrow source-frame spectral windows of $(2650-2680)${\AA} and $(3010-3070)${\AA} in which the Fe~II template of \cite{2001ApJS..134....1V} shows significant local minima.
This continuum is subtracted from the observed spectra, and the residual is integrated over the wavelength range of $(2650-3070)${\AA} to approximate the Mg~II + Fe~II flux.
Individual spectra are then stacked with weights proportional to the exposure time of each observation.

The quality of individual residual spectra is evaluated by the rms function calculated over the continuum fitting spectral windows.
These rms values were found to be in the range of $(1-25)\times 10^{-17}\;{\rm erg\,s^{-1}\,cm^{-2}\,\AA^{-1}}$.
After analysing the strongest outlier values among estimates of the Mg~II + Fe~II line flux, we decided to introduce an upper limit to the rms value: ${\rm rms}_{\rm max} = 7\times 10^{-17}\;{\rm erg\,s^{-1}\,cm^{-2}\,\AA^{-1}}$. Applying this criterion, we rejected 107 (19\%) of the most noisy individual photometric spectra from further analysis. This selection was also applied to the results of spectral decomposition analysis.

\subsection{Fermi-LAT}

Gamma-ray data from the Fermi Large Area Telescope \citep{2009ApJ...697.1071A} were extracted from the region of interest (ROIs) of radius $10^\circ$ centred on the position of 3C~454.3.
We used the {\tt ScienceTools} software package, version {\tt v11r5p3}, to perform the maximum likelihood analysis, fitting a {\tt PowerLaw2} spectral model over the energy range $100\;{\rm MeV}$ -- $100\;{\rm GeV}$, using the {\tt P8R2\_SOURCE\_V6} response function, maximum zenith angle of $100^\circ$, and a source model including all background sources from the 3FGL catalogue \citep{2015ApJS..218...23A} up to the angular separation of $25^\circ$, with the detection criteria of ${\rm TS} > 10$ and $N_{\rm pred} > 3$.\footnote{We adopt a less strict detection criterion than the usual ${\rm TS} > 25$ in order to cover the period of the lowest gamma-ray flux at ${\rm MJD} = 55800-56200$.}
The light curve was obtained over the period of ${\rm MJD} \simeq 54683 - 58234$ with uniform time bins of 10 days.

\section{Results}
\label{sec_res}

Figure \ref{fig_lc} shows $\gamma$-ray and optical light curves of 3C~454.3 over the period of 10 years (2008 - 2018; MJD 54700 - 58300). The $\gamma$-ray data show large-amplitude (factor $\sim 100$) variations that can be broadly characterised as two periods of high activity (2009.5-2011.0 and 2013.5-2018) separated by a particularly quiescent period of 2011.0-2013.5.
Due to the relatively long binning  timescale (10 days), short-scale variations (detected down to the timescale of $\simeq 5\;{\rm h}$; \citealt{2017Galax...5..100N}) with peak fluxes reaching the level of $f(E>100\;{\rm MeV}) \simeq 8\times 10^{-5}\;{\rm ph\,s^{-1}\,cm^{-2}}$ on MJD~55520 \citep{2011ApJ...733L..26A} are not included here.
However, taking them into account, the overall variability amplitude exceeds a factor of 1000.

\begin{figure}
\includegraphics[width=\columnwidth]{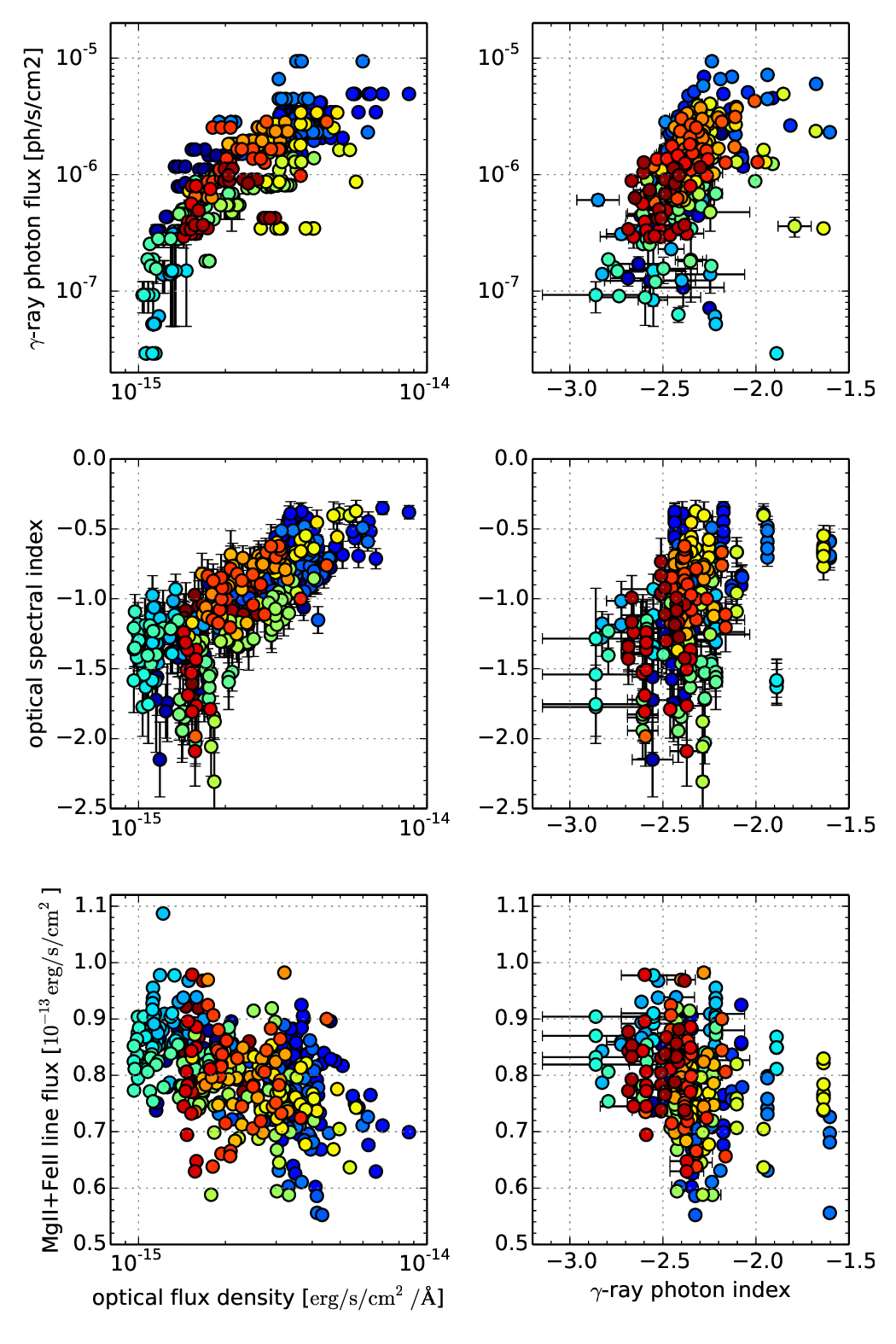}
\caption{Scatter diagrams between $\gamma$-ray photon flux, $\gamma$-ray photon index, optical continuum flux density, optical spectral index, and integrated Mg~II + Fe~II line flux measured simultaneously within $\pm 5$ days. The colour-coding indicates the observation date, as in the top panel of Figure \ref{fig_lc}.}
\label{fig_scatter}
\end{figure}

Simultaneous variations in the optical continuum are roughly correlated with the $\gamma$-ray signal {(cf. \citealt{2009ApJ...697L..81B,2012AJ....143...23G,2017MNRAS.464.2046K,2017MNRAS.472..788G})}, but with significantly lower amplitude (a factor of $\sim 20$). It has been noticed before that the $\gamma$-ray flux scales in at least quadratic relation to the optical continuum \citep{2011MNRAS.410..368B}. This is confirmed by the scatter diagram of $\gamma$-ray photon flux versus optical flux density shown in the top left panel of Figure \ref{fig_scatter}. In the lowest gamma-ray state at MJD~$\sim 56000$, for $\gamma$-ray fluxes $< 10^{-7}\;{\rm ph\,s^{-1}\,cm^{-2}}$, the optical $\lambda_{\rm src} = 3000\AA$ continuum flux density does not decrease below $10^{-15}\;{\rm erg\,s^{-1}\,cm^{-2}\,\AA}$, or the level at which contribution of the thermal accretion disc emission is expected to become dominant. The optical spectral index $\alpha$ (defined as $f_\lambda \propto \lambda^\alpha$) is generally lower during the low $\gamma$-ray and optical state (see the middle left panel of Figure \ref{fig_scatter}), which means a redder-when-brighter trend consistent with the presence of accretion disc component  \citep[also known as the big blue bump, e.g.][]{2006A&A...450...39G,2006A&A...453..817V,2007A&A...473..819R,2009ApJ...697L..81B}, although the value of $\alpha$ remains highly variable.

We present two different estimates of the optical continuum; one is based on the spectral decomposition (blue circles in Figure \ref{fig_lc}) including the spectral template for Fe~II emission and the other is a simplified approach (red crosses in Figure \ref{fig_lc}) in which Fe~II emission is not separated from the Mg~II emission. The normalisations of the optical continuum at $\lambda_{\rm src} = 3000\AA$ and the spectral indices $\alpha$ obtained with these two methods are consistent. On the other hand, the fluxes of the broad Mg~II emission line integrated over $\lambda_{\rm obs}$ appear discrepant in the fourth panel of Figure \ref{fig_lc} {because of the inclusion of Fe II in simplified measurements}. However, for the initial period (${\rm MJD} < 55800$) we compare them with the values for Mg~II and Mg~II+Fe~II obtained from the same Steward spectra and reported in  Table 1 in \cite{2013ApJ...763L..36L}\footnote{The $(1+z)^3$ K-correction was applied to the line fluxes reported by \cite{2013ApJ...763L..36L}; here its effect is removed for the green symbols shown in Figure \ref{fig_lc}.}
We find that our decomposition results are consistent with the Mg~II fluxes of \cite{2013ApJ...763L..36L}, and that our simplified results are consistent with their Mg~II+Fe~II fluxes.
On the other hand, the Mg~II flux measurements based on the SMARTS observations, as reported by \cite{2013ApJ...779..100I,2015ApJ...804....7I}, are systematically lower (a  factor of $\sim 3$) than the Mg~II fluxes obtained from the Steward data. A comparison of these two datasets is discussed in \cite{2013ApJ...779..100I,2015ApJ...804....7I}; however, the relative spectral calibration between these two instruments has not been fully explained.

The Mg~II line flux light curve reveals several outliers with respect to the typical values, and occasional differences with the results of \cite{2013ApJ...763L..36L}.
Observation at MJD~54771 yields both the line fluxes (either decomposed Mg~II or simplified Mg~II+Fe~II) and the continuum level that are a factor of $\sim 3$ higher than several other observations taken within the same week (with consistent line FWHM and continuum index); a corresponding systematic difference is also seen in the photometric spectrum.
A similar outlier can be seen at MJD~56308, with continuum and line fluxes increased by factor $\sim 1.5$;  unlike the case of MJD~54771, this spectrum is affected by noise despite the same exposure time ($960\;{\rm s}$).
Several observations around MJD~55295 suggest line fluxes lower than the typical values, and also lower than the results of \cite{2013ApJ...763L..36L};  these observations are characterised by relatively short exposure times ($320\;{\rm s}$), and hence high statistical flux uncertainty.

\subsection{Short-term variations}

\cite{2013ApJ...763L..36L} reported an episode of short-term variation (flare) in the Mg~II line flux (and in the Fe~II emission) during an optical flare centred at MJD~55512.
This epoch is indicated in Figure \ref{fig_lc} by the green dotted lines.
We confirm that both Mg~II and Mg~II+Fe~II line fluxes are elevated during the week-long window centred at MJD~55512,  compared with neighbouring windows.
Because of wide gaps in the observations, the variability timescale is uncertain: it could be closer to weekly than daily.
However, all individual spectra obtained over that period are relatively noisy;  their rms values are above our adopted upper limit ${\rm rms}_{\rm max}$ (see  bottom panel of Figure \ref{fig_lc}).
We searched for more examples of such flares beyond MJD~55900, and we have not found any. Around MJD~56832, there was an even brighter flare of the optical continuum \citep{2017MNRAS.464.2046K}, but it was not accompanied by any clear increase in the Mg~II line flux (and the individual spectra obtained over that week are even more noisy than those around MJD~55512).

\begin{figure}
\includegraphics[width=\columnwidth]{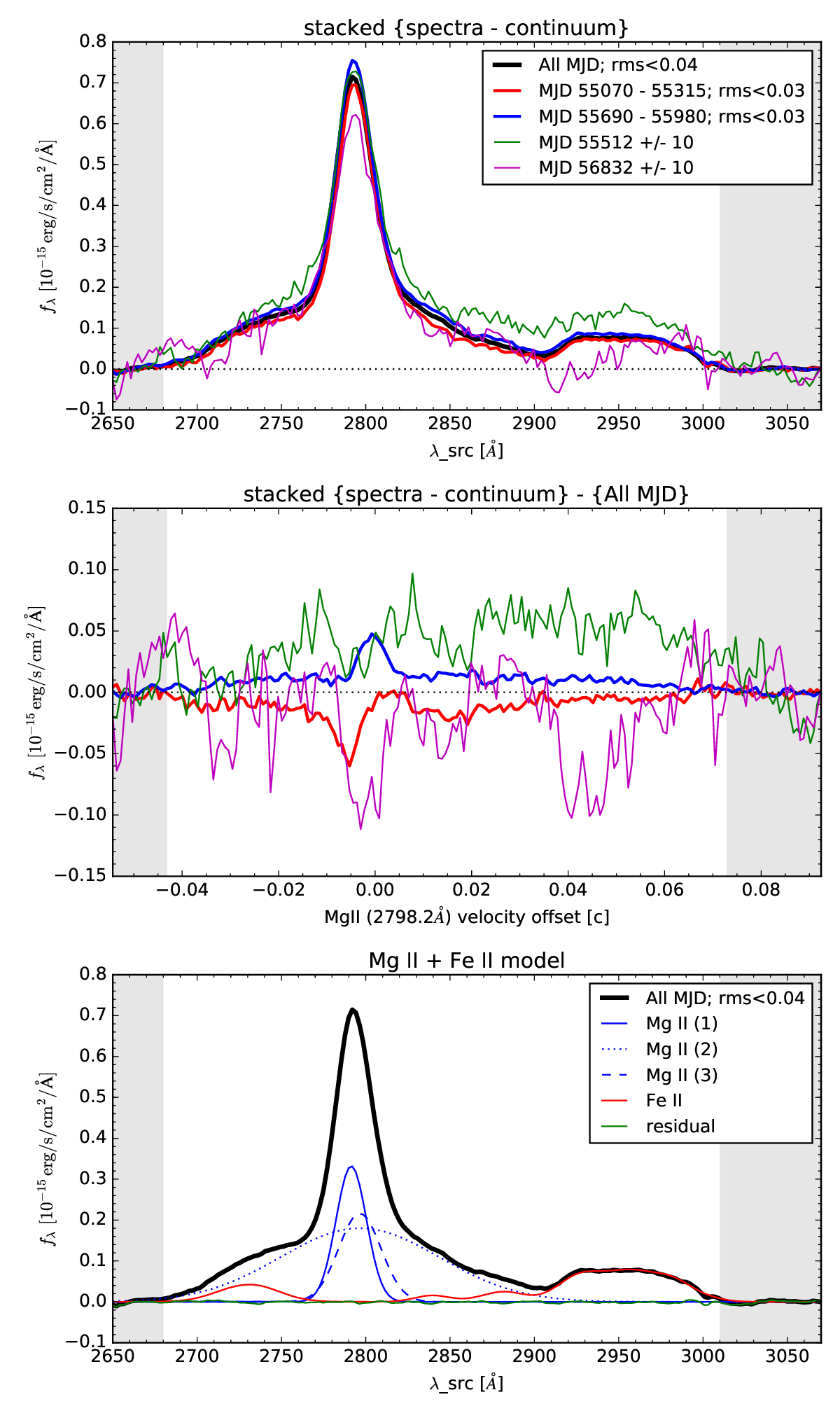}
\caption{Upper panel: Observed source-frame spectra stacked over different observational periods after subtraction of power-law continuum. Middle panel: Differences between spectra stacked over limited observational periods and the spectrum stacked over all available observations (line styles correspond to those in the upper panel). Lower panel: model for the spectral decomposition of the Mg~II line and theoretical Fe~II pseudo-continuum template. The spectral windows for continuum evaluation are indicated with grey stripes.}
\label{fig_stack}
\end{figure}

In Figure \ref{fig_stack}, we present stacked continuum-subtracted spectra evaluated over two short observational windows (without any selection for the rms values): ${\rm MJD} = 55512 \pm 10$ (green line) and ${\rm MJD} = 56832 \pm 10$ (magenta line). These line profiles are compared with a ten-year average\footnote{Based on 273 individual spectra characterised by ${\rm rms} < 3\times 10^{-17}\;{\rm erg\,s^{-1}\,cm^{-2}\,\AA^{-1}}$ calculated over the continuum fitting spectral windows.} (thick black line). Differences between short-term stacked spectra and the ten-year average are shown in the lower panel of Figure \ref{fig_stack}. We performed two-sample Kolmogorov-Smirnov (K-S) tests for the difference between stacked spectra.
This test suggests that the line profile stacked during the MJD~55512 flare reported by \cite{2013ApJ...763L..36L} is clearly different from the ten-year average ($p \sim 10^{-9}$; $6\,\sigma$), while the line profile stacked during the second optical continuum flare around MJD~56832 is only different at the level of $2\,\sigma$ ($p \simeq 0.04$).

\begin{figure}
\includegraphics[width=\columnwidth]{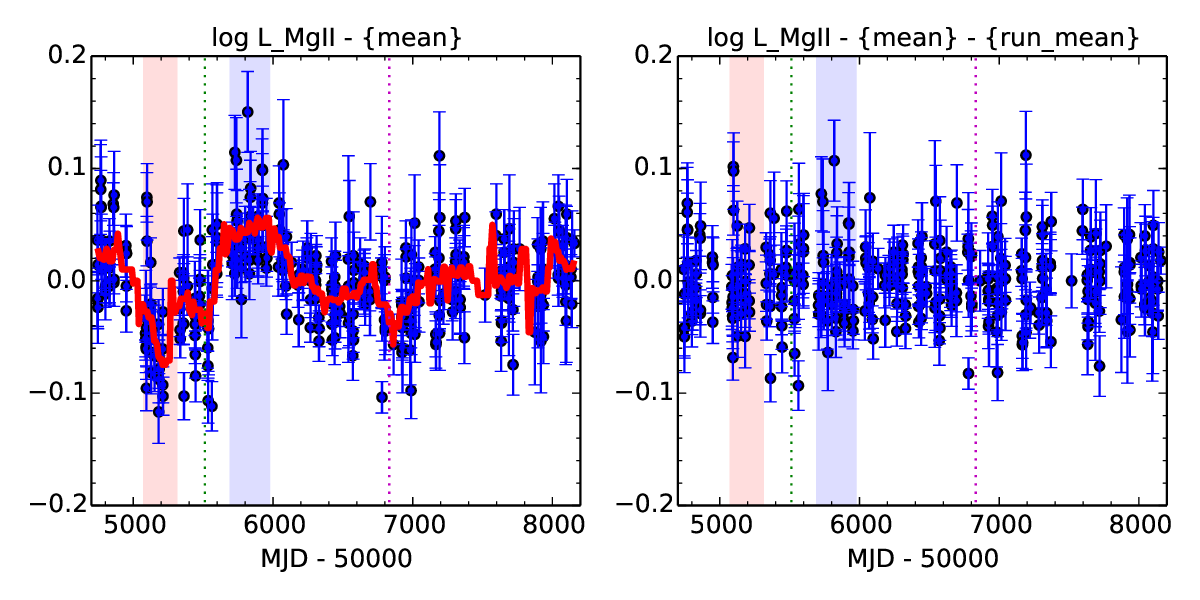}
\caption{Left panel:  Light curve of Mg~II line luminosity $\log_{10} L_{\rm MgII}$ obtained with the spectral decomposition method, with the mean value subtracted (blue);  running mean calculated for the minimum timescale of $100\;{\rm d}$ (red). Right panel: Residual of the light curve shown in the left panel with the running mean subtracted. Periods of interest for spectral stacking are marked with vertical stripes or lines (cf. Figure \ref{fig_lc}).}
\label{fig_runmean}
\end{figure}

\subsection{Long-term variations}

The left panel of Figure \ref{fig_runmean} shows again the light curve of Mg~II line luminosity obtained from the spectral decomposition method (corresponding to the blue circles shown in the fourth panel of Figure \ref{fig_lc}, we take the logarithm and subtract the mean value).
Due to our selection for ${\rm rms} < {\rm rms}_{\rm max}$, the short-term variations discussed in the previous subsection are largely excluded.
We suggest that this light curve shows a systematic modulation on long timescales ($\sim 100\;{\rm days)}$.
{A power spectral density (PSD) analysis showed that variations in the Mg~II line luminosity are consistent with the red noise, with no sign of quasi-periodic oscillations (QPOs).}

\begin{figure}
\centering
\includegraphics[width=0.8\columnwidth]{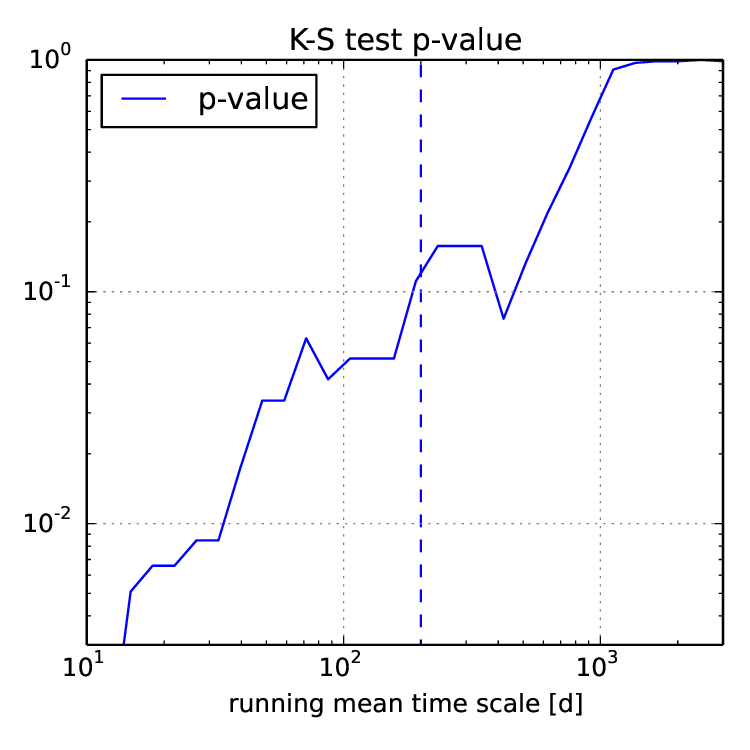}
\caption{Probability of rejecting the hypothesis that subtracting a running mean does not affect the distribution of Mg~II line luminosities calculated from two-sample K-S  tests between   the measured light curve of $\log L_{\rm MgII}$ (left panel of Figure \ref{fig_runmean}) and the running mean subtracted light curve (right panel of Figure \ref{fig_runmean}) as function of the running mean timescale.}
\label{fig_kstest}
\end{figure}

In order to evaluate the significance of this long-term modulation, we calculate a running mean light curve using uniform time bins of fixed timescale $\tau$. As an example, in the left panel of Figure \ref{fig_runmean}, we show the running mean light curve for $\tau = 100\;{\rm d}$ (solid red line), and in the right panel of that figure we show the measured Mg~II luminosities after subtracting the running mean. We then perform a series of   two-sample K-S tests between the measured and running mean subtracted Mg~II light curves for a wide range of running mean timescales $\tau$. Figure \ref{fig_kstest} shows the dependence of the $p$-values for the null hypothesis (that subtracting a running mean does not modify the distribution of Mg~II luminosities) on the running mean timescale $\tau$. The basic result is that $p < 0.1$ for $\tau < 200\;{\rm d}$, hence long-term variations have only moderate statistical significance ($1.5\,\sigma$).

\begin{figure}
\includegraphics[width=\columnwidth]{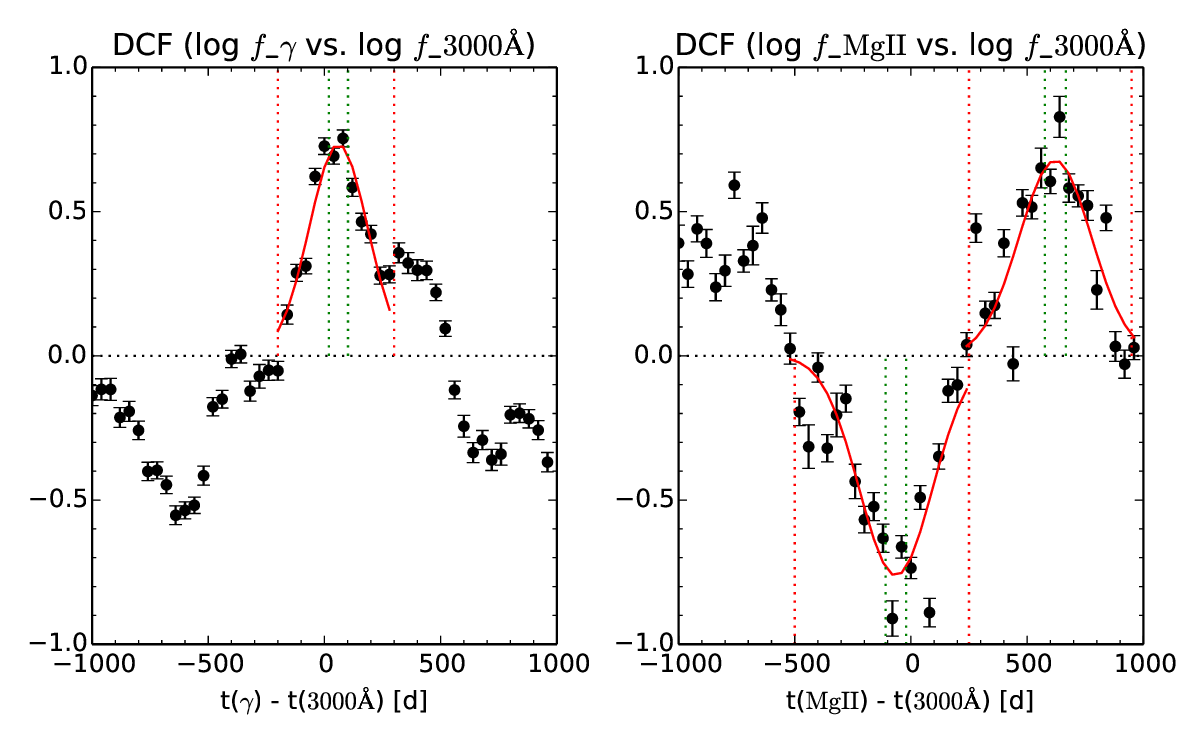}
\caption{Discrete correlation functions calculated between the logarithms of the ($>100\;{\rm MeV}$) $\gamma$-ray photon flux and the optical continuum flux density (left panel), and between the logarithms of the Mg~II line flux and the optical continuum (right panel).}
\label{fig_dcf}
\end{figure}

We also calculated the discrete correlation functions (DCF; \citealt{1988ApJ...333..646E}) in order to estimate the characteristic time lags between partially correlated signals.\footnote{Very similar results were obtained with the ZDCF algorithm of \cite{1997ASSL..218..163A}.} Noting the fundamentally different nature of variations in the Mg~II line luminosity (moderate departures from the mean value) and the optical continuum (large-amplitude--order of magnitude variations without a clear mean value), we cross-correlated the logarithms of measured flux values using the time lag bins of 40~days. As a reference, we first calculated the DCF between the $\gamma$-ray and optical continua (left panel of Figure \ref{fig_dcf}). This DCF has a clear positive peak of ${\rm DCF} \simeq 0.73$ at $\Delta t \simeq 0$. In order to estimate the most likely value of time lag, we fitted a Gaussian function over a limited range of time lags (red dotted lines), finding a centroid of $\Delta t_0 = 60\pm 41\,{\rm d}$ and dispersion of $\sigma(\Delta t) = 178\pm 42\;{\rm d}$.
The right panel of Figure \ref{fig_dcf} shows the DCF calculated between the Mg~II line luminosity and the optical $\lambda_{\rm src} = 3000\AA$ continuum. This DCF shows broad negative and positive peaks of ${\rm DCF} \simeq -0.76$ and $0.68$ at $\Delta t \simeq 0$ and $\Delta t \simeq 600\;{\rm d}$, respectively.
A Gaussian fit to the negative peak yields the centroid of $\Delta t_0 = -65\pm 44\,{\rm d}$ and dispersion of $\sigma(\Delta t) = 222\pm 48\;{\rm d}$;
a fit to the positive peak yields the centroid of $\Delta t_0 = 621\pm 45\,{\rm d}$ and dispersion of $\sigma(\Delta t) = 221\pm 50\;{\rm d}$.
This allows for two alternative interpretations: (1) that the Mg~II luminosity is anti-correlated with the optical continuum (as can be   seen in the bottom left panel of Figure \ref{fig_scatter}), or (2) that the Mg~II luminosity lags behind the optical continuum by $\Delta t \simeq 600\;{\rm days}$. The two interpretations are discussed in Section \ref{sec_dis}.

Looking at the Mg~II light curve more subjectively, we can distinguish two periods of interest: a period of reduced Mg~II luminosity at ${\rm MJD} \simeq 55070 - 55315$ and a period of enhanced Mg~II luminosity at ${\rm MJD} \simeq 55690 - 55980$ (see Figures \ref{fig_lc} and \ref{fig_runmean}). The midpoints of these periods are separated by 640 days. We use these periods to calculate another continuum-subtracted stacked spectra presented in Figure \ref{fig_stack}. From a two-sample K-S test between these two stacked spectra,
we find a probability $p \simeq 5\times 10^{-4}$ (confidence level of $3.5\,\sigma$) that these spectra are consistent with each other.

\subsection{Modelling the mean line profile}

In Figure \ref{fig_stack}, we   show   an example of  an accurate model for the decomposition of the stacked continuum-subtracted spectrum into an Mg~II line represented by three Gaussian components \citep[e.g.][]{2018ApJ...856...78A} with the FWHM of $1607, 8390$, and $2239\;{\rm km\,s^{-1}}$, and a theoretical Fe~II pseudo-continuum with independent multiplets of 60-63, 78, and I~Zw~1 \citep{2015ApJS..221...35K,2019MNRAS.484.3180P} with kinematic broadenings in the range $2400 - 3600\;{\rm km\,s^{-1}}$.

\section{Discussion}
\label{sec_dis}

In many non-blazar AGNs, variations in the BEL flux are observed to be correlated with the optical continuum (typically interpreted as accretion disc emission) with time delays of $\Delta t \sim (10-1000)\;{\rm days}$ \citep[e.g.][]{2000ApJ...533..631K,2004ApJ...613..682P,2007ApJ...659..997K}. Measuring this time delay allows us to estimate the characteristic radius $R_{\rm BLR}$ of the BLR; this technique is known as reverberation mapping \citep{1982ApJ...255..419B}.

\begin{figure}
\includegraphics[width=\columnwidth]{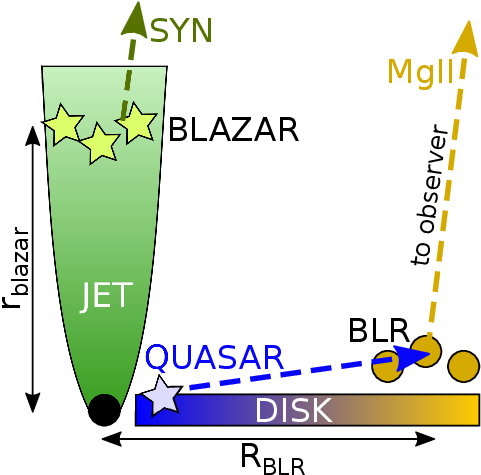}
\caption{Schematic illustration of an  AGN  being both a quasar (luminous thermal emission from the accretion disc) and a blazar (even more luminous non-thermal emission from the relativistic jet), i.e. an FSRQ like 3C~454.3. It is assumed here that the BLR is concentrated along the accretion disc plane at characteristic distance $R_{\rm BLR}$ from the SMBH.
The optical continuum is dominated by the synchrotron radiation produced in a blazar zone located within the jet at characteristic distance $r_{\rm blazar}$, and the Mg~II BEL is produced in the BLR.}
\label{fig_cartoon}
\end{figure}

In the case of 3C~454.3, which is both a blazar and a quasar, we have a more complex situation (see Figure \ref{fig_cartoon}). The optical continuum is a combination of highly variable synchrotron emission produced in a blazar zone located in a relativistic jet up to a parsec in front of the nucleus (SMBH), and of thermal emission produced in the accretion disc. It is not possible to confidently separate these two components. In the traditional picture of BLR being concentrated and slightly flattened around the equatorial plane of the AGN \citep[e.g.][]{2012arXiv1209.2291T,2014ApJ...789..161N,2015MNRAS.449..431J,2018Natur.563..657G}, it is this unknown accretion disc emission (rather than the jet emission that is relativistically beamed away from the equatorial plane) that is reprocessed into the BEL. Observation of a correlation between synchrotron light curve produced in a foreground jet and BEL light curve produced in the background BLR requires that both are correlated with the underlying poorly constrained accretion disc emission. While the reverberation of accretion disc emission in the BLR can be explained purely by light-travel effects, correlation of the synchrotron emission requires consideration of perturbations originating in the inner accretion disc and propagating along the jet to the blazar zone.

{We should note that the DCF presented in the right panel of Figure \ref{fig_dcf}, between the Mg~II line luminosity and the optical continuum, shows a broad negative peak around zero time lag, suggesting that these two light curves could be anti-correlated. This anti-correlation would be difficult to explain in the traditional models of BLR geometry, such as the one depicted in Figure \ref{fig_cartoon}. It could be possibly accommodated in a variant of the extended BLR scenario, with the BLR clouds concentrated along the relativistic jets and ionised by their relativistically boosted non-thermal radiation, as advocated by \cite{2013ApJ...763L..36L}. Instead of having short-scale flares of line luminosity, positively correlated with the optical continuum flares, increasing luminosity of jet emission could have a destructive effect on such extended BLRs by inducing too strong ionisation of the surrounding gas. The main difficulty of such a scenario would be to explain a much longer variability timescale ($\sim$ 1 month) of the BEL.}

Assuming that the observed time delay of $\Delta t_{\rm obs} \simeq (621\pm 45)\;{\rm days}$ between the optical continuum light curve and the Mg~II light curve is due to reverberation of the thermal accretion disc emission, the corresponding additional path distance that should be travelled by reverberated light in the source frame is $\Delta r = c\Delta t_{\rm obs}/(1+z) \simeq (0.28 \pm 0.02)\;{\rm pc}$.
For a blazar observer located at small viewing angle $\theta_{\rm obs} \sim \Gamma_{\rm j}^{-1}$ measured from the jet axis, the relative distance measured along the line of sight between the blazar zone and the BLR zone is $r_{\rm blazar}\cos\theta_{\rm obs}$, where $r_{\rm blazar} \sim 0.1-1\;{\rm pc}$ \citep{2014ApJ...789..161N} is the distance along the jet of the synchrotron emitting blazar zone.
Considering that at $t = 0$, a perturbation in the inner accretion disc produces a flare in the quasar continuum emission, and triggers a perturbation (e.g. internal shock) propagating along the jet with velocity $\left<\beta_j\right> = \left<v_j\right>/c$. The continuum flare illuminates the BLR zone at $ct_{\rm BLR} \simeq R_{\rm BLR}$, while the jet perturbation reaches the blazar zone at $ct_{\rm blazar} \simeq r_{\rm blazar}/\left<\beta_j\right>$. We can thus express the additional reverberation path distance as $\Delta r = r_{\rm blazar}\cos\theta_{\rm obs} + ct_{\rm BLR} - ct_{\rm blazar} = R_{\rm BLR} - r_{\rm blazar}\left(\left<\beta_j\right>^{-1} - \cos\theta_{\rm obs}\right)$.
With this, we can attempt to constrain the BLR radius:
\begin{equation}
R_{\rm BLR} \simeq \Delta r + \left(\left<\beta_j\right>^{-1} - \cos\theta_{\rm obs}\right)r_{\rm blazar} \gtrsim \Delta r\,.
\label{eq_RBLR}
\end{equation}
Since $\left<\beta_j\right>^{-1} \gtrsim 1 \gtrsim \cos\theta_{\rm obs}$ and $r_{\rm blazar} \sim \Delta r$, the second term in the above equation is not important.
This corresponds to an independent estimate of the black hole mass,
\begin{equation}
\frac{M_{\rm SMBH}}{M_\odot} = f\left(\frac{R_{\rm BLR}}{R_{\rm g,\odot}}\right)\left(\frac{v_{\rm rms}}{c}\right)^2 \gtrsim (8.5 \pm 2.3)\frac{f}{1.3}\times 10^8\,,
\end{equation}
where
$v_{\rm rms} \simeq (3170 \pm 410)\;{\rm km\,s^{-1}}$
is measured from the mean FWHM of the Mg~II line and $f \sim 1.3$ is a geometrical form factor \citep{2018Natur.563..657G}.
The obtained value is very reasonable and corresponds well to the previous estimates $5-45\times 10^8 M_\odot$, based mostly on the empirical relation between line widths and continuum luminosities \citep{2001MNRAS.327.1111G, 2002ApJ...579..530W, 2006ApJ...637..669L, 2011MNRAS.410..368B, 2012MNRAS.421.1764S, 2017MNRAS.472..788G}.
In the case of a blazar with the jet oriented close to the line of sight, a BLR that is concentrated along the AGN equatorial plane should be oriented close to the plane of the sky, largely eliminating the uncertainty on the BLR inclination angle.

{Our claim for delayed correlation between optical continuum and broad line luminosity is based in the first place on the modulation of the line luminosity as presented in Figure \ref{fig_runmean}.
While we identified two periods of interest in the line luminosity light curve, their connection to the structure of the continuum light curve is not obvious.
The period of enhanced line luminosity at ${\rm MJD} \simeq 55690-55980$ could be related to the continuum flare peaking at ${\rm MJD} \simeq 55150$ (corresponding to a time delay of $\sim 690$ days) or to the following continuum flare peaking at ${\rm MJD} \simeq 55520$ (with a delay of $\sim 320$ days).
On the other hand, the period of reduced line luminosity at ${\rm MJD} \simeq 55070-55315$ could be related to the continuum dip at ${\rm MJD} \simeq 54900$ (with delay of $\sim 290$ days), while a longer delay of $600-700$ days would point to the period of ${\rm MJD} \simeq 54500-54600$, which coincides with a seasonal gap in the optical photometric data from the GASP-WEBT project \citep{2010ApJ...712..405V}.
Finally, the historical minimum in the continuum light curve at ${\rm MJD} \sim 56000$ does not have a corresponding decrease in line luminosity, which we would expect at ${\rm MJD} \sim 56600-56700$. Unlike the case of 3C~273 \citep{2019ApJ...876...49Z}, this tentative reverberation measurement should be treated with caution, and should also be confirmed by future spectroscopic monitoring campaigns.

\begin{acknowledgement}
We thank the referee for helpful suggestions, and Marek Sikora and Bo\.zena Czerny for discussions.
We used data from the Steward Observatory spectropolarimetric monitoring project that has been supported by Fermi Guest Investigator grants NNX08AW56G, NNX09AU10G, NNX12AO93G, and NNX15AU81G.
We also used public data acquired by the Fermi Large Area Telescope created by NASA and DoE (USA) in collaboration with institutions from France, Italy, Japan, and Sweden.
KN was supported by the Polish National Science Centre grant 2015/18/E/ST9/00580.
The work of ACG is partially supported by Indo-Poland project no. DST/INT/POL/P19/2016 funded by the Department of Science and Technology (DST) of the Government of India, and by the Chinese Academy of Sciences (CAS) President's International Fellowship Initiative (PIFI) grant no. 2016VMB073.
KH acknowledges the support of the Polish National Science Centre grant 2015/17/B/ST9/03422.
MFG was supported by the National Science Foundation of China (grants 11873073 and U1531245).
ML was supported by the National Science Foundation of China (grants 11773056 and U1831138).
\end{acknowledgement}

\end{document}